\RequirePackage{ifpdf}
\ifpdf 
\documentclass[pdftex]{sigma}
\else
\documentclass{sigma}
\fi

\def \Tr {\mathop{\rm Tr}\nolimits}

\newcommand \widebar [1] {\overline{#1}}

\newcommand \vev [1] {\langle{#1}\rangle}

\newcommand{\as}{\ifmmode\alpha_{\rm s}\else{$\alpha_{\rm s}$}\fi}
\newcommand{\asbar}{\ifmmode\bar{\alpha}_{\rm s}\else{$\bar{\alpha}_{\rm s}$}\fi}

\newcommand \CR {\mathcal{R}}

\newcommand \VV {\mathbb{V}}

\newcommand{\balpha} {{\boldsymbol{\alpha}}}
\newcommand{\bbeta} {{\boldsymbol{\beta}}}

\newcommand{\bv}{{\boldsymbol{v}}}

\font\cmss=cmss12 
\def\inbar{\,\vrule height1.5ex width.4pt depth0pt}
\def\IC{\relax\hbox{$\inbar\kern-.3em{\rm C}$}}
\def\IZ{\relax{\hbox{\cmss Z\kern-.4em Z}}}
\def\IR{{\hbox{{\rm I}\kern-.2em\hbox{\rm R}}}}

\def\IP{{\hbox{{\rm I}\kern-.2em\hbox{\rm P}}}}
\def\II{\hbox{{1}\kern-.25em\hbox{l}}}

\newbox\lett\newdimen\lheight\newdimen\lwidth
\def\ontop#1#2{\setbox\lett=\hbox{#2}\lheight\ht\lett
\multiply\lheight by 12 \divide\lheight by 10\relax%
\lwidth\wd\lett \multiply\lwidth by 8 \divide\lwidth by 10\relax #2\kern-\lwidth%
\raise\lheight\hbox{{$\scriptstyle #1$}}\kern.1ex}

\def\inbar{\,\vrule height1.5ex width.4pt depth0pt}

\numberwithin{equation}{section}

\begin{document}
\renewcommand{\PaperNumber}{084}

\FirstPageHeading

\renewcommand{\thefootnote}{$\star$}

\ShortArticleName{$\CR$-Matrix and Baxter $\mathcal{Q}$-Operators}

\ArticleName{$\boldsymbol{\CR}$-Matrix and Baxter
$\boldsymbol{\mathcal{Q}}$-Operators\\ for the Noncompact
$\boldsymbol{{SL(N,\mathbb{C})}}$ Invariant Spin
Chain\footnote{This paper is a contribution to the Vadim Kuznetsov
Memorial Issue ``Integrable Systems and Related Topics''. The full
collection is available at
\href{http://www.emis.de/journals/SIGMA/kuznetsov.html}{http://www.emis.de/journals/SIGMA/kuznetsov.html}}}

\Author{Sergey \'E. DERKACHOV~$^\dag$ and Alexander N.
MANASHOV~$^{\ddag\S}$}

\AuthorNameForHeading{S.\'E.~Derkachov and A.N.~Manashov}

\Address{$^\dag$~St.-Petersburg Department of Steklov
Mathematical Institute of Russian Academy of Sciences,\\
$\phantom{^\dag}$~Fontanka 27, 191023 St.-Petersburg, Russia}

\EmailD{\href{mailto:derkach@euclid.pdmi.ras.ru}{derkach@euclid.pdmi.ras.ru}}
\Address{$^\ddag$~Institute for Theoretical Physics, University of
Regensburg, D-93040 Regensburg, Germany}
\EmailD{\href{mailto:alexander.manashov@physik.uni-regensburg.de}{alexander.manashov@physik.uni-regensburg.de}}

\Address{$^\S$~Department of Theoretical Physics,
Sankt-Petersburg  University, St.-Petersburg, Russia}

\ArticleDates{Received October 30, 2006; Published online December 02, 2006}

\Abstract{The problem of constructing   the $SL(N,\mathbb{C})$
invariant solutions to the Yang--Baxter equation is considered.
The solutions ($\CR$-operators) for arbitrarily principal series
representations of  $SL(N,\mathbb{C})$ are obtained in an explicit
form. We construct the commutative family of the operators
$\mathcal{Q}_k(u)$ which can be identif\/ied with the Baxter
operators for the noncompact $SL(N,\mathbb{C})$ spin magnet.}


\Classification{82B23; 82B20}

\begin{flushright}
\it To the memory of Vadim Kuznetsov
\end{flushright}

\section{Introduction}

\looseness=1 
The Yang--Baxter equation (YBE) plays an important
role  in the theory of completely integrable systems. Its
solutions, the so-called $\CR$-matrices (operators), are  basic
ingredients of Quantum Inverse Scattering Method
(QISM)~\cite{FST,LF}. The problem of constructing solutions to the
YBE was thoroughly analyzed in the works of
Drinfeld~\cite{Drinfeld,Drinfeld2}, Jimbo~\cite{Jimbo} and many
others. In the most studied case of  the quantum af\/f\/ine Lie
(super)algebras the universal $\CR$-matrix was constructed in
works of Rosso~\cite{Rosso}, Kirillov and Reshetikhin~\cite{KR}
and Khoroshkin and Tolstoy~\cite{KT1,KT2}. The interrelation  of
YBE with the representation theory was elucidated by Kulish,
Reshetikhin and Sklyanin~\cite{KRS,KR82}, who studied solutions of
YBE  for f\/inite dimensional representations of the
$GL(N,\mathbb{C})$ group. The $\CR$-matrix on  inf\/inite
dimensional spaces were not considered till recently.

\renewcommand{\thefootnote}{\arabic{footnote}}
\setcounter{footnote}{0}

\looseness=1 
In the present paper we construct the
$SL(N,\mathbb{C})$ invariant
 $\mathcal{R}$-operator which acts on the tensor product of two principal series
representations of the $SL(N,\mathbb{C})$ group. The spin chains
with an inf\/inite dimensional Hilbert space, so called noncompact
magnets, are interesting in connection with the problem of
constructing the Baxter $\mathcal{Q}$-operators~\cite{Baxter} and
the representation of Separated Variables
(SoV)~\cite{Sklyanin84,Sklyanin}. The Baxter
$\mathcal{Q}$-operators are known now for a~number of models.
Beside few exceptions these are the spin chains with a symmetry
group of  rank one. No regular method of constructing Baxter
operators for models with symmetry groups of higher rank exists so
far. However, in the studies of the noncompact $sl(2)$ magnets it
was noticed that the transfer matrices with generic (inf\/inite
dimensional) auxiliary space are factori\-zed into the product of
Baxter operators. This property seems to be  quite a general
feature of the noncompact magnets\footnote{This property holds
also for the integrable models with $U_q(\widehat{sl}_{n})$
symmetry, at least for $n=2,3$~\cite{BLZ,BHK}.} and can be related
to the factorization of $\CR$-matrix suggested in~\cite{SD}. Thus
the problem of constructing the Baxter operators is reduced, at
least for the noncompact spin chains, to the problem of
factorization of the $\CR$-matrix. The factorizing operators for
the
 $sl(N)$ invariant $\CR$-matrix acting
on the tensor product of two generic lowest weight $sl(N)$ modules
for $N=2,3$ were constructed in~\cite{SD}. Unfortunately, for a
higher $N$ the def\/ining equations for the factorizing operators
become too complicated to be solved directly.

To get some insight into a possible structure of
solutions for a general $N$ it is instructive  to consider the
problem of constructing an $\CR$-operator for  principal series
representations of $SL(N,\mathbb{C})$. We  remark here that
contrary to naive expectations  the principal series noncompact
magnets appear to be in some respects simpler than their
(in)f\/inite dimensional cousins. Dif\/ferent spin chain models
with $sl(N)$ symmetry  dif\/fer  by a functional realization of a
quantum space. The less restrictions are imposed on the functions
from a quantum space, the simpler becomes the analysis of
algebraic properties of a model and the harder its exact solution.
For instance, the solution of the noncompact $SL(2,\mathbb{C})$
spin magnet~\cite{SL2C,DKKM} presents a quite nontrivial problem
already for the spin chain of the length $L=3$, while the analysis
of this model (constructing the Baxter operators, SoV
representation, etc) becomes considerably easier in comparison
with its compact analogs. In particularly, the $\CR$-operator in
this model ($SL(2,\mathbb{C})$ spin magnet) has a rather simple
form  and, as one can easily verify, admit the factorized
representation. All this suggests to consider the problem of
constructing the $\CR$-operator for the principal series
representations of $SL(N,\mathbb{C})$ in f\/irst instance. In the
paper we give the complete solution to this problem. We will
obtain the explicit  expression for the $\CR$-operator, prove that
the latter satisf\/ies Yang--Baxter equation and, as was expected,
enjoys the factorization property. It allows us to construct the
Baxter $\mathcal{Q}$-operators as  traces of a product of the
factorizing operators.

We hope that the obtained results will also be  useful for the
construction of the representation of Separated Variables. So far
the latter  was constructed in an explicit form only for a very
limited number of models (see
\cite{KNS,KL,KLS,SL2C,DKM-I,DKM-II,KMS,S,Iorgov}). It was noticed
by Kuznetsov and Sklyanin~\cite{KS,Sklyanin00} that the kernel of
a separating operator and  Baxter operator have to be related to
each other. So the knowledge of the Baxter operator for the
$SL(N,\mathbb{C})$ magnets can shed light on the form of the
separating operator.

The paper is organized as follows. In   Section~\ref{ind} we
recall the basic facts about the principal series representations
of the complex unimodular group. We construct the operators which
intertwine the equivalent representations and describe their
properties. 
This material is well known \cite{Knapp1,Knapp2} but we  represent
intertwinning operators in the form which is appropriate for our
purposes.
Section~\ref{roperator} starts with the description of
our approach for constructing the $\CR$-operator. We derive the
def\/ining equations on the  $\CR$-operator and solve them.
Finally, we show that the obtained operator satisf\/ies the
Yang--Baxter equation. In  Section~\ref{sect:fq} we study the
properties of the factorizing operators in more detail and discuss
their relation with the Baxter $\mathcal{Q}$-operators. Finally we
construct the explicit realization of the operators in question as
integral operators.

\section[Principal series representations of the group $SL(N,\mathbb{C})$]{Principal
series representations of the group
$\boldsymbol{SL(N,\mathbb{C})}$} \label{ind}

The unitary principal series representations of the group
$SL(N,\mathbb{C})$ can be constructed as
follows~\cite{Gelfand,SG}. Let $Z$ and $H$  be  the groups of the
lower triangular matrices with unit elements   on a diagonal and
the upper triangular matrices with unit determinant, respectively,
\begin{gather*}
z=
\begin{pmatrix}1&0&0&\ldots&0\\
z_{21}&1&0&\ldots&0\\
z_{31}&z_{32}&1&\ldots&0\\
\vdots&\vdots&\vdots&\ddots&\vdots\\
z_{N1}&z_{N2}&\ldots&z_{N,N-1}&1
\end{pmatrix}\in Z,\qquad
h=
\begin{pmatrix}
  h_{11} & h_{12} & h_{13}&\ldots & h_{1N} \\
  0 & h_{22} & h_{23}&\ldots & h_{2N} \\
  0 & 0 & h_{33}&\ldots & h_{3N} \\
  \vdots & \vdots & \vdots&\ddots &\vdots \\
  0 & 0 &0 &\ldots & h_{N,N}
\end{pmatrix}\in H.
\end{gather*}
Almost any  matrix ${g}\in G=SL(N,\mathbb{C})$ admits the Gauss
decomposition ${g}=z h$ (the group $Z$ is the right coset $G/H$).
The element $z_1\in Z$ satisfying the condition
\begin{gather}\label{bzh}
g^{-1}\cdot z=z_1\cdot h
\end{gather}
will be denoted by $z\bar g$, so that $g^{-1}z=z\bar g\cdot h$.
Thus we can speak about a local right action of $G$ on $Z$. Later
in \eqref{Tg}, also $h$, dependent on $g$ and $z$ as
in~\eqref{bzh}, will be used again.

Let $\balpha$  be a character of the group $H$ def\/ined by the
formula
\begin{gather}\label{dh}
{\balpha}(h)=\prod_{k=1}^N h_{kk}^{-\sigma_k-k}\,\bar
h_{kk}^{-\bar\sigma_k-k},
\end{gather}
where $\bar h_{kk}\equiv (h_{kk})^*$ is the complex conjugate of
$h_{kk}$, whereas in general  $\sigma_k^*\neq \bar \sigma_k$.
Since $\det h=1$ the function $\balpha(h)$ depends only on the
dif\/ferences $\sigma_{k,k+1}\equiv\sigma_k-\sigma_{k+1}$   and
can be rewritten in the form
\begin{gather*}
\balpha(h)=\prod_{k=1}^{N-1} \Delta_{k}^{1-\sigma_{k,k+1}} \bar
\Delta_{k}^{1-\bar\sigma_{k,k+1}}
=\prod_{k=1}^{N-1}\Delta_{k}^{n_k} \, \vert
\Delta_{k}\vert^{2(1-\bar\sigma_{k,k+1})},
\end{gather*}
where $\Delta_{k}=\prod\limits_{i=1}^k h_{ii}$ and $n_k= \bar
\sigma_{k,k+1}-\sigma_{k,k+1}$, $k=0,\ldots,N-1$,
 are  integer numbers\footnote{From now in, since each
variable $a$ comes along with its  antiholomorphic twin $\bar a$
we will write down only holomorphic variant of equations.}. One
can always assume that  parameters $\sigma_k$ satisfy the
restriction
\begin{gather}\label{rs}
\sigma_1+\sigma_2+\cdots+\sigma_N=N(N-1)/2.
\end{gather}
The map $g\to T^\alpha(g)$, where
\begin{gather}\label{Tg}
[T^{\balpha}(g)\Phi](z)=\balpha(h^{-1})\Phi(z\bar g),
\end{gather}
def\/ines a principal series representation of the group
$SL(N,\mathbb{C})$ on a suitable space of functions on the group
$Z$,
$\Phi(z)=\Phi(z_{21},z_{31},\ldots,z_{NN-1})$~\cite{Gelfand,SG}.
The operator $T^\alpha(g)$ is a unitary operator on the Hilbert
space $L^2(Z)$,
\begin{gather*}
\vev{\Phi_1|\Phi_2}=\int \prod_{1\leq i<k\leq N} d^2 z_{ki}\,
(\Phi_1(z))^*\, \Phi_2(z)\,,
\end{gather*}
if the character   $\balpha'(h)=\balpha(h)\prod\limits_{k=1}^N
|h_{kk}|^{2k}$ is a unitary one, i.e.\ $\vert\balpha'\vert=1$.
This condition  holds if $\sigma_{k,k+1}^*+\bar\sigma_{k,k+1}=0$
for $k=1,\ldots,N-1$, i.e.
\begin{gather}\label{unitarity}
\sigma_{k,k+1}=-\frac{n_k}2+i\lambda_k, \qquad
\bar\sigma_{k,k+1}=\frac{n_k}2+i\lambda_k,\qquad k=1,2,\ldots,N-1,
\end{gather}
$n_k$ is integer and $\lambda_k$ is real.

The unitary principal series representation $T^{\balpha}$ is
irreducible. Two representations
 $T^\balpha$ and  $T^{\balpha'}$ are unitary equivalent if and only if the
corresponding parameters $(\sigma_1,\ldots,\sigma_N)$ and
$(\sigma'_1,\ldots,\sigma'_N)$ are related to each other by a
permutation~\cite{Gelfand,SG}.

\subsection{Generators and  right shifts}
We will need the explicit expression for the generators of
inf\/initesimal $SL(N,\mathbb{C})$ transformations. The latter are
def\/ined in the standard way
\begin{gather*}
\left[T^\balpha\left(\II+\sum_{ik}\epsilon^{ik}
\mathcal{E}_{ki}\right)\Phi\right](z)=
\Phi(z)+\sum_{ik}\epsilon^{ik}E_{ki} \Phi(z)+O(\epsilon^2),
\end{gather*}
where $\mathcal{E}_{ik}$, ($1\leq i,k\leq N$), are the generators
in the fundamental representation of the  $SL(N,\mathbb{C})$
group,
\begin{gather*}
(\mathcal{E}_{ik})_{nm}=\delta_{in}\delta_{km}-\frac1N\delta_{ik}\delta_{nm}.
\end{gather*}
The generators $E_{ik}$ are  linear dif\/ferential operators in
the variables $z_{mn}$, ($1\leq n<m\leq N$) which satisfy the
commutation relation
\begin{gather}\label{comm}
[E_{ki} ,E_{nm}]=\delta_{in} E_{km}-\delta_{km} E_{ni}.
\end{gather}
It follows from def\/inition~\eqref{Tg} that the lowering
generators $E_{ki}$, $k>i$, are the generators of  left shifts,
$\Phi(z)\to L({z_0}) \Phi(z)=\Phi(z_0^{-1} z)$. In a similar way
we def\/ine the  generators
 of  right shifts,
$\Phi(z)\to R({z_0}) \Phi(z)=\Phi(z z_0)$,
\begin{gather}\label{rshifts}
\Phi\left(z\left(\II+\sum_{k>i}\epsilon^{ik}\mathcal{E}_{ki}\right)\right)=
\left(1+\sum_{k>i}\epsilon^{ik}D_{ki}+O\big(\epsilon^2\big)\right)\Phi(z).
\end{gather}
Since  right and left shifts commute one concludes that $[E_{ki},
D_{nm}]=0$ ($k>i, n>m$). Clearly, the generators $D_{ki}$ satisfy
the same commutation relation as $E_{ki}$,~equation~\eqref{comm}
\begin{gather*}
[D_{ki} ,D_{nm}]=\delta_{in} D_{km}-\delta_{km} D_{ni}.
\end{gather*}
The explicit expression for the generators of left and right
shifts reads
\begin{gather}\label{ED}
E_{ki}=-\sum_{m=1}^{i}
z_{im}\,\frac{\partial\phantom{z_{m}}}{\partial z_{km}}, \qquad
D_{ki}=\sum_{m=k}^N
z_{mk}\,\frac{\partial\phantom{z_{m}}}{\partial z_{mi}}=
-\sum_{m=1}^{i} \tilde
z_{im}\frac{\partial\phantom{z_{m}}}{\partial{\tilde z_{km}}},
\end{gather}
where $\tilde z_{ki}=(z^{-1})_{ki}$ and  we recall that
$z_{ii}=1$. Let us notice here that the operator $D_{ki}$ depends
on the variables in the $k$-th and $i$-th columns of the matrix
$z$, or  on the variables in the  $k$-th and $i$-th rows of the
inverse matrix $z^{-1}$.

The generators $E_{ki}$ can be expressed in terms of the
generators $D_{ki}$ as follows
\begin{gather}\label{DE}
E_{ki}=-\sum_{mn}z_{im}\bigl(D_{nm}+\delta_{nm}\sigma_m\bigr)\big(z^{-1}\big)_{nk},
\end{gather}
where $D_{nm}$ is nonzero only for $n>m$. To derive
equation~\eqref{DE} it is suf\/f\/icient to notice that for the
inf\/initesimal transformation
 $g=\II+\epsilon\cdot \mathcal{E}=\II+\sum_{ik}\epsilon^{ik}\mathcal{E}_{ki}$ the elements of the
 Gauss decomposition of the matrix $g^{-1}z$ can be represented in the
 following form:
 $z\bar g=z(\II-\epsilon_-(z))$ and $h=\II-\epsilon_+(z)$. The matrices
 $\epsilon_-(z)$ and $\epsilon_+(z)$ are the lower and upper diagonal parts
 of the matrix $z^{-1}(\epsilon\cdot \mathcal{E})z$. Namely, the matrix $\epsilon_-(z)_{ki}$ is
 nonzero only for $k>i$,
$\epsilon_-(z) _{ki}=(z^{-1}(\epsilon \cdot \mathcal{E}) z)_{ki}$,
while $\epsilon_+(z) _{ki}=(z^{-1}(\epsilon \cdot \mathcal{E})
z)_{ki}$, for $k\leq i$ and zero otherwise. Thus one f\/inds that
the inf\/initesimal transformation~\eqref{Tg} is essentially the
right shift generated by $z_0=\II-\epsilon_-(z)$. Finally, taking
into account that $\sum\sigma_k=N(N-1)/2$ one obtains the
representation~\eqref{DE} for the symmetry generators.

It is convenient to rewrite equation~\eqref{DE} in the matrix form
\begin{gather}\label{DM}
E
=-z\left(D+\sigma\right) z^{-1},
\end{gather}
where $E
=\sum_{mn}e_{mn}E_{nm}$, $D=\sum_{n>m}e_{mn} D_{nm}$ and
$\sigma=\sum_{n}e_{nn}\sigma_n$ and the matrices $e_{nm}$ form the
standard basis in $\textrm{Mat}(N\times N)$,
$(e_{nm})_{ik}=\delta_{in}\delta_{mk}$. Let us recall that $z$ and
$z^{-1}$ are lower triangular matrices while $D+\sigma$ is an
upper triangular matrix
\begin{gather}\label{Ds}
D+\sigma=\begin{pmatrix}\sigma_1 &D_{21}&D_{31}&\ldots&D_{N1}\\
0&\sigma_2& D_{32}&\ldots& D_{N2}\\
\vdots&\vdots& \ddots& \vdots&\vdots\\
0&0&\ldots&\sigma_{N-1}&D_{N,N-1}\\
0&0&\ldots&0&\sigma_{N}
 \end{pmatrix}
\end{gather}
The dependence of the
generators $E
$ on the representation $T^\balpha$ resides in the parameters
$\sigma_n$ entering  the matrix $\sigma$. The same formulae hold
for the antiholomorphic generators $\bar E$.

Closing this subsection we give  some  identities that will be
useful in the further analysis:
\begin{subequations}\label{Dc}
\begin{gather}
D_{ki} z=z\,e_{ki},\\
D_{ki}\, z^{-1}=-e_{ki}   z^{-1},\\
\label{DDik} [D_{ki},
D]=\sum_{n<i}D_{kn}e_{ni}-\sum_{n>k}e_{kn}D_{ni}.
\end{gather}
\end{subequations}
The f\/irst two formulae follow directly from
def\/inition~\eqref{rshifts}, while the last one is a consequence
of the commutation relations for the generators $D_{ki}$. We
remind also that the operators $D_{ki}$ are def\/ined (nonzero)
only for $k>i$.

\subsection{Intertwining operators}\label{twin}
It is known that  two  principal series representations
$T^\balpha$ and $T^{\balpha'}$ are equivalent if the parameters
which  specify the representations,
$\vec{\sigma}=(\sigma_1,\ldots,\sigma_N)$ and
$\vec{\sigma}'=({\sigma}'_1,\ldots,\sigma'_N)$, are related to
each other by a permutation, $\vec{\sigma}'=P\,\vec{\sigma}$ (of
course, it is assumed that the parameters $\{\bar \sigma_k \}$,
$\{\bar \sigma'_k \}$ in the antiholomorphic sector are related by
the same permutation, $\vec{\bar\sigma}'=P\vec{\bar
\sigma}$)~\cite{Gelfand}. An arbitrary permutation can be
represented as a composition of the elementary permutations,
$P_{k}$, which interchange the $k$-th and $k+1$-st component of
the vectors, $\vec{\sigma}\, (\vec{\bar\sigma})$,
\begin{align*}
P_{k},(\ldots,\sigma_k,\sigma_{k+1},\ldots)=(\ldots,\sigma_{k+1},\sigma_k,\ldots).
\end{align*}
The operator $U_k$ intertwining the representations, $T^\balpha$
and $T^{\balpha'}$ which dif\/fer  by the elementary permutation
of spins $P_{k}$, $\balpha'=P_{k}\balpha$,~  has the form
\begin{gather}   \label{U-twin}
 U_k=D_{k+1,k}^{\sigma_{k,k+1}}\bar D_{k+1,k}^{\bar \sigma_{k,k+1}},\qquad
U_k T^\balpha=  T^{\balpha'} U_k,
\end{gather}
where $\sigma_{k,k+1}=\sigma_k-\sigma_{k+1}$ ($\bar
\sigma_{k,k+1}=\bar \sigma_k-\bar \sigma_{k+1}$), and $D_{k+1,k}
(\bar{D}_{k+1,k})$ is  the generator of the right shift. Let us
note that the operator $D_{k+1,k}$ depends on the variables
$z_{nm}$ in the $k$-th and $k+1$-st columns of the matrix $z$.
After the change of variables
\begin{gather}\label{ztox}
z_{k+1,k}=x_{k+1,k},\quad z_{m,k+1}=x_{m,k+1},\quad
z_{m,k}=x_{m,k}+ x_{k+1,k}x_{m,k+1},\quad k<m\leq N
\end{gather}
the operator $D_{k+1,k}$ turns into a derivative with respect to
$x_{k+1,k}$, $D_{k+1,k}=\partial_{x_{k+1,k}}$, which means that
the construction given in equation~\eqref{U-twin} results in
well-def\/ined operator. Moreover, if the powers
in~equation~\eqref{U-twin} satisfy the condition $
\sigma_{k,k+1}^*+\bar\sigma_{k,k+1}=0$, the operator $U_k$ is a
unitary operator on $L^2(Z)$.

To prove that operator $U_k$ is an intertwining  operator it is
suf\/f\/icient to check that it intertwines the generators $E$ and
$E^\prime$ in the representations $T^{\balpha}$ and
$T^{\balpha'}$, $E^{{\prime}} U_k=U_k  E$. To this end it is
convenient to use the representation~\eqref{DM} for the
generators. Using commutation relations~\eqref{Dc} one f\/inds
\begin{gather}\label{Dcc}
U_k z=z\left(1+\alpha D_{k+1,k}^{-1} e_{k+1,k}\right) U_k ,\qquad
U_k z^{-1}=\left(1-\alpha D_{k+1,k}^{-1} e_{k+1,k}\right)
z^{-1}U_k
\end{gather}
and
\begin{gather}\label{Dccc}
U_k D= \left(D+\alpha
D_{k+1,k}^{-1}\left(\sum_{n<k}D_{k+1,n}e_{nk}-\sum_{n>k+1}e_{k+1,n}
D_{n,k}\right)\right) U_k,
\end{gather}
where $\alpha=\sigma_{k,k+1}$. (Let us note that the operator
$D_{k+1,k}^{-1}$ commutes with $e_{k+1,k}z^{-1}$ and with the
operators in the sum in~\eqref{Dccc}, so its position can be
changed). Starting from $U_k E$ and moving the operator $U_k$ to
the right with the help of equations~\eqref{Dcc} and \eqref{Dccc}
one gets after some simplif\/ications
\begin{gather*}
U_k E= -z\left(D+\sigma+\alpha(e_{k+1,k+1}-e_{k,k})-
\alpha(\alpha-\sigma_{k,k+1})e_{k+1,k} D^{-1}_{k+1,k}\right)z^{-1} U_k\nonumber\\
\phantom{U_k E}{}=-z(D+\sigma')z^{-1}U_k=E^\prime U_k.
\end{gather*}

(One has to take into account that $\alpha=\sigma_{k,k+1}$ and the
matrix $\sigma^\prime=\sigma+\sigma_{k,k+1} (e_{k+1,k+1}-e_{k,k})$
dif\/fers from the matrix $\sigma$ by the transposition
$\sigma_{k}\leftrightarrow \sigma_{k+1}$.)

Making use of the change of variables~\eqref{ztox} one can
represent the operator $U_k$ in the form of an integral operator
\begin{gather}\label{prop}
\lbrack U_k\Phi\rbrack(z)= A(\sigma_{k,k+1})\int d^2\zeta
{[z_{k+1,k}-\zeta]^{-1-\sigma_{k,k+1}}} \Phi(z_\zeta),
\end{gather}
where $[z]^{\sigma}=z^{\sigma} \bar z^{\bar\sigma}$,
 \begin{gather}  \label{A-factor}
A(\sigma)\overset{\mathrm{def}}{=}A(\sigma,\bar\sigma)= \frac1\pi
{i^{\bar\sigma-\sigma}} {\Gamma(1+\sigma)}/{\Gamma(-\bar\sigma)}
\end{gather}
and
\begin{gather*}
z_\zeta=z\bigl(1+(\zeta-z_{k+1,k}) e_{k+1,k}\bigr).
\end{gather*}
Let us note that the matrices $z$ and $z_\zeta$ dif\/fer from each
other by the elements in the  $k$-th column only.

The operator $U_k$ depends on the dif\/ference of
$\sigma_{k,k+1}$,
 $U_k=U(\sigma_{k,k+1})$, which is
determined by the representation which the operator acts on. We do
not  display the dependence on this parameter explicitly, but want
to note here  that in the product $U_{k+1}U_k$, the  f\/irst
operator is $U_{k}=U_k(\sigma_{k,k+1})$,  while the second one is
$U_{k+1}= U_{k+1}(\sigma_{k,k+2})$, since the operator $U_k$
interchanges the parameter $\sigma_k$ with $\sigma_{k+1}$.
Similarly, one f\/inds that $U_k\, U_k=U_k(\sigma_{k+1,k})
U_{k}(\sigma_{k,k+1})=\II$.

 The intertwining operators $U_k$ satisfy the same commutation
relations as the operators of elementary permutations $P_{k}$
\begin{subequations}\label{UUU}
\begin{alignat}{3}
&U_k U_k=\II, &&& \\
& U_kU_n=U_k\,U_n, \qquad\qquad&&  \text{for}\quad |k-n|>1&\\
& U_k U_{k+1}U_k=U_{k+1} U_{k}U_{k+1}.&&&
 \end{alignat}
\end{subequations}
The f\/irst relation had been already explained. The second one is
a trivial consequence of the commutativity of the generators of
right shifts, $[D_{k+1,k},D_{n+1,n}]=0$ for $|k-n|>1$. The last
relation can be checked by the direct calculation with the help
of the integral representation~\eqref{prop}.

Obviously, the operator intertwining the representations
$T^\balpha$ and $ T^{\balpha'}$  such that the characters
$\balpha$ and $\balpha'$  are related to each other by some
permutation can be constructed as a certain combination of the
operators $U_k$, $k=1,\ldots,N-1$. The intertwining operators,
$U_k$, play an important role in the construction of an
$\mathcal{R}$-operator which we will discuss in the next section.

\section[$\mathcal{R}$-operator]{$\boldsymbol{\mathcal{R}}$-operator}\label{roperator}

Let us recall that an $\mathcal{R}$-operator is a linear operator
which acts on the tensor product of two spaces $\VV_1\otimes
\VV_2$, depends on the spectral parameter $u$ and satisf\/ies the
Yang--Baxter relation
\begin{gather}   \label{YB}
\CR_{12}(u-v)\CR_{13}(u-w)\CR_{23}(v-w)=\CR_{23}(v-w)\CR_{13}(u-w)\CR_{12}(u-w).
\end{gather}
As usual, it is implied that the operator $\CR_{ik}(u)$ acts
nontrivially on the tensor product $\VV_i\otimes \VV_k$. In the
case under consideration each space $\VV_i$ ($\VV_k$) is assumed
to be  a vector space of some  representation of the group
$SL(N,\mathbb{C})$. We are interested in constructing an
$\CR$-operator which acts on the space $L^2(Z)\otimes L^2(Z)$ and
is invariant with respect to $SL(N,\mathbb{C})$ transformations,
\begin{gather*}
[T^{\balpha_1}(g)\otimes T^{\balpha_2}(g), \CR_{12}(u)]=0.
\end{gather*}
The form of  the $\CR$-operator strongly depends  on the spaces
$\VV_1$ and  $\VV_2$ which it acts on. The $\CR$-operator has the
simplest form  when both space $\VV_1$ and $\VV_2$ is the vector
space of  the  fundamental ($N$-dimensional) representation of the
$SL(N,\mathbb{C})$ group, $\VV_1=\VV_2=\VV_f$. Namely, in the case
that $\VV_1=\VV_2=\VV_3=\VV_f$ the solution of the Yang--Baxter
equation~\eqref{YB} is given by the operator
$\CR_{ik}(u)=u+P_{ik}$, where $P_{ik}$ is the permutation operator
on $\VV_i\otimes \VV_k$. This solution can also be represented in
the  form
\begin{gather*}
\CR_{ik}(u)=u+\sum_{mn}e_i^{nm}e_k^{mn},
\end{gather*}
Substituting  the generators $e_{i}^{nm}$ by the generators
$E^{nm}$ in some generic representation of the $SL(N,\mathbb{C})$
group in the above formula one gets the $\CR$-operator on the
space $\VV_1\otimes \VV_f$. Such an operator is called Lax
operator. Let us recall here that in a generic representation of
the $SL(N,\mathbb{C})$ group there are two sets of generators,
holomorphic  $E_{ik}$ and the antiholomorphic $\bar{E}_{ik}$, so
that we def\/ine two Lax operators, the holomorphic one, $L(u)$,
and the antiholomorphic one, $\bar L(\bar u)$,
\begin{gather}\label{Lax}
L(u)=u+\sum_{mn} e_{mn}\, {E}_{nm}, \qquad \bar L(\bar u)=\bar
u+\sum_{mn} e_{mn} \bar{E}_{nm}.
\end{gather}
The Yang--Baxter equation~\eqref{YB} on the spaces
$\VV_1\otimes\VV_2\otimes\VV_f$ takes the form
\begin{subequations}\label{R-LL}
\begin{gather}
\CR_{12}(u-v,\bar u-\bar v)L_1(u)L_2(v) =
L_2(v)L_1(u)\CR_{12}(u-v,\bar u-\bar v),\\
\CR_{12}(u-v,\bar u-\bar v)\bar L_1(\bar u)\bar L_2(\bar v) = \bar
L_2(\bar v)\bar L_1(\bar u)\CR_{12}(u-v,\bar u-\bar v).
\end{gather}
\end{subequations}
These equations can be considered as def\/ining equations for the
operator $\CR_{12}(u,\bar u)$. Let us notice that the
$\CR$-operator depends on two spectral parameters $u$ and $\bar
u$, which are not supposed to be related each other. We will show
later  that the spectral parameters are subject to the
restriction, $u-\bar u=n$, where $n$ is an integer
number\footnote{The quantization of the spectral parameters for
the $SL(2,\mathbb{C})$ spin magnet was observed in~\cite{SL2C}.}.
In the next subsection we discuss the approach for solving
equations~\eqref{R-LL} suggested in~\cite{SD}.

\subsection[Factorized ansatz for $\CR$-matrix]{Factorized ansatz for $\boldsymbol{\CR}$-matrix} \label{sect:FR}

Let us remark  that  the Lax operator~\eqref{Lax} depends on the
spectral parameter $u$ and $N$ parame\-ters~$\sigma_{k}$. Since
the character $\balpha$~(equation~\eqref{dh}) and, hence,
 the generators of the $SL(N,\mathbb{C})$ group
 depend only on the dif\/ferences $\sigma_{k,k+1}=\sigma_k-\sigma_{k+1}$,
one concludes that the Lax operators depend on $N$-independent
parameters which can be chosen as $u_k=u-\sigma_k$,
$k=1,\ldots,N$, $L(u)=L(u_1,\ldots,u_N)$.
 These parameters appear quite naturally when one uses the
representation~\eqref{DM} for the generators. Indeed, the  Lax
operator can be represented as
\begin{gather}\label{Lfact}
L(u)=z\big(u-\sigma-D\big)z^{-1},
\end{gather}
and the parameters $u_k$ are nothing else as diagonal elements of
the matrix $(u-\sigma-D)$.

Let us represent the operator $\CR_{12}$ in the form
$\CR_{12}=P_{12}\mathbb{R}_{12}$. Note, that since
$\VV_1=\VV_2=L^2(Z)$ the permutation operator $P_{12}$ is
unambiguously def\/ined on $\VV_1\otimes \VV_2$. We will associate
the matrix variables $z$ and $w$ with the spaces $\VV_1$ and
$\VV_2$, respectively,  so that the vectors in $\VV_1\otimes
\VV_2$ are functions $\Phi(z,w)$. The action of the permutation
operator $P_{12}$ is def\/ined in a conventional way,
$P_{12}\Phi(z,w)=\Phi(w,z)$. The Yang--Baxter
relation~\eqref{R-LL} can be rewritten as follows\footnote{From
now on we will write down the equations in the holomorphic sector
only and suppress  the dependence on ``barred'' variables, $\bar
u,\bar v$, etc for brevity.}
\begin{gather}
\mathbb{R}_{12}(u-v)L_1(u_1,\ldots,u_N)L_2(v_1,\ldots,v_N)\nonumber\\
\qquad{}=
L_1(v_1,\ldots,v_N)L_2(u_1,\ldots,u_N)\mathbb{R}_{12}(u-v).\label{RLL-1}
\end{gather}
The parameters $v_k$ are def\/ined as $v_k=v-\rho_k$ where the
parameters $\{\rho\}$  specify
 the representation of $SL(N,\mathbb{C})$ group on  the space $\VV_2$.
It is seen from equation~\eqref{RLL-1} that  action of the
operator $\mathbb{R}_{12}(u-v)$ results in permutation of the
arguments $\{u\}$ and $\{v\}$ of the Lax operators~$L_1$
and~$L_2$. Since we have already  constructed
 the operators $\{U_k\}$ which interchange the components of the string
$\{u\}$ ($\{v\}$) it seems reasonable to try to f\/ind  operators
that carry out permutations of the strings $\{u\}$ and $\{v\}$.

It was suggested in \cite{SD} to look for  the operator
$\mathbb{R}_{12}$ in the factorized form
\begin{gather} \label{RRk}
\mathbb{R}_{12}(u-v)=\mathbb{R}^{(1)}\,\mathbb{R}^{(2)}\cdots\mathbb{R}^{(N)},
\end{gather}
where each operator $\mathbb{R}^{(k)}$ interchanges the  arguments
$u_k$ and $v_k$ of the Lax operators,
\begin{gather}
\mathbb{R}^{(k)}L_1(u_1,\ldots, u_k,\ldots u_N)L_2(v_1,\ldots,v_k,\ldots,v_N)\nonumber\\
\qquad{}{}=L_1(u_1,\ldots,v_k,\ldots,u_N)L_2(v_1,\ldots,u_k,\ldots,v_N)\mathbb{R}^{(k)}.\label{RkL}
\end{gather}
Evidently, if such operators can be constructed then
equation~\eqref{RLL-1} will follow immediately from
equations~\eqref{RRk} and \eqref{RkL}. We note also that since any
permutation inside the string $\{u\}$ can be carried out with the
use of the operators  $U_k$ (and similarly for the string
$\{v\}$), it is suf\/f\/icient to f\/ind  an operator which
interchanges some elements $u_k$ and $ v_i$ of strings $\{u\}$ and
$\{v\}$.

\subsection{Exchange operator}\label{exchange}
In this subsection we will construct operator $S$ which exchanges
the arguments $u_1$ and $v_N$ of the Lax operators $L_1$ and
$L_2$,
\begin{gather}
{S}\,L_1(u_1,u_2,\ldots, u_N)L_2(v_1,\ldots,v_{N-1},v_N)\nonumber\\
\qquad{}=
L_1(v_N,u_2,\ldots,u_N)L_2(v_1,\ldots,v_{N-1},u_1)S.\label{SLL}
\end{gather}
It turns out that  the operator $S$ has a surprisingly simple
form. Taking into account that the character
$\balpha(h)$~(equation~\eqref{dh}) can be written in the form
\begin{gather}  \label{au}
\balpha(h)=\prod_{k=1}^N h_{kk}^{u_k-k}\bar h_{kk}^{\bar u_k-k},
\end{gather}
it is easy to f\/igure out that operator $S$ has to intertwine the
representations $T^\balpha\otimes T^\bbeta$ and
$T^{\balpha'}\otimes T^{\bbeta'}$,
\begin{gather}\label{ST}
S\big(T^\balpha\otimes T^\bbeta\big)=\big(T^{\balpha'}\otimes
T^{\bbeta'}\big)S,
\end{gather}
where the  characters $\balpha$ and $\balpha'$, ($\bbeta$ and
$\bbeta'$) are related  to each  other as follows (see
equation~\eqref{dh})
\begin{gather*}
{\balpha}'(h)=h_{11}^{v_N-u_1}\bar h_{11}^{\bar v_N-\bar u_1}
\balpha(h), \qquad \bbeta'(h)=h_{NN}^{u_1-v_N} \bar h_{NN}^{\bar
u_1-\bar v_N} \bbeta(h).
\end{gather*}
It turns out that the simplest operator intertwining the
representations in question gives  a solution to
equation~\eqref{SLL}. To construct an operator with the required
transformation properties let us consider the matrix $w^{-1} z$.
Under the transformation $z\to g^{-1}z=(z\bar g) h(z,g)$,  $w\to
g^{-1}w=(w\bar g)h(w,g)$ it transforms as follows
\begin{gather*}
w^{-1}z=h^{-1}(w,g)(w\bar g)^{-1}(z\bar g) h(z,g).
\end{gather*}
Taking into account that the matrices $z$ and $w$ are  lower
triangular matrices while $h(z,g)$, $h(w,g)$ are upper triangular
ones, one f\/inds that the matrix element $(w^{-1}z)_{N1}$
transforms in a simple way
\begin{gather}   \label{wzt}
\big(w^{-1}z\big)_{N1}=h^{-1}_{NN}(w,g) \big((w\bar g)^{-1}(z\bar
g)\big)_{N1}h_{11}(z,g).
\end{gather}
It suggests to def\/ine the operator $S$ as follows
\begin{gather}   \label{defS}
\left[S(\gamma,\bar
\gamma)\Phi\right](z,w)=\big(\big(w^{-1}z\big)_{N1}\big)^\gamma
\big({\big(w^{-1}z\big)}_{N1}^*\big)^{\bar \gamma} \Phi(z,w),
\end{gather}
where $\gamma=u_1-v_N$, $\bar\gamma=\bar u_1-\bar v_N$. Notice,
that the dif\/ference $\gamma-\bar\gamma$ has to be an integer
number. \vskip 3mm It follows from equations~\eqref{wzt} and
\eqref{defS} that the operator $S$ has necessary transformation
properties~\eqref{ST}. Thus it remains  to show that the operator
$S$ satisf\/ies equation~\eqref{SLL}. We start with two useful
identities for the Lax operator $L(u_1,\ldots,u_N)$
\begin{gather} \label{zLz}
\sum_m z_{Nm}^{-1} L_{mk}=u_N\,z_{Nk}^{-1},\qquad \sum_m L_{km}
z_{m1}=u_1 z_{k1},
\end{gather}
which follow immediately from the representation~\eqref{Lfact} and
a triangularity of the matrix $D$, equation~\eqref{Ds}. Next,
using the def\/initions of the operator of right
shifts~\eqref{rshifts},~\eqref{ED} one f\/inds
\begin{gather*}
\left(\sum_{k>i} e_{ik} D_{ki}^{(z)}\right)
\big(w^{-1}z\big)_{N1}=\sum_{k>1} e_{1k} \big(w^{-1} z\big)_{Nk}=
-\big(w^{-1}z\big)_{N1} e_{11}+\sum_{k\geq 1} e_{1k} \big(w^{-1}
z\big)_{Nk}.
\end{gather*}
Then it is straightforward to derive
\begin{gather}\label{SL1}
S(\gamma,\bar \gamma) \left(L_1\right)_{nm} S^{-1}(\gamma,\bar
\gamma) = \big(z(u-\gamma e_{11} -D) z^{-1}\big)_{nm}+
\frac{\gamma}{(w^{-1}z)_{N1}} z_{n1} w^{-1}_{Nm}.
\end{gather}
where $u$ is a diagonal matrix, $u=\sum_k u_k e_{kk}$. Taking into
account that $\gamma=u_1-v_N$ one f\/inds that the f\/irst term in
the rhs is the Lax operator $L(v_N,u_2,\ldots,u_N)$. Quite
similarly, one derives for the Lax operator $L_2$
\begin{gather}
S(\gamma,\bar \gamma) \left(L_2\right)_{nm}
S^{-1}(\gamma,\bar\gamma) = \big(w (v+\gamma e_{NN} -D)
w^{-1}\big)_{nm}-
\frac{\gamma}{(w^{-1}z)_{N1}} z_{n1} w^{-1}_{Nm}\nonumber\\
\phantom{S(\gamma,\bar \gamma) \left(L_2\right)_{nm}
S^{-1}(\gamma,\bar\gamma)}{}
=\big(L_2(v_1,\ldots,v_{N-1},u_1)\big)_{nm}-\frac{\gamma}{(w^{-1}z)_{N1}}
z_{n1} w^{-1}_{Nm}.\label{SL2}
\end{gather}
where $v=\sum_k v_k e_{kk}$. The equation~\eqref{SLL} then follows
immediately from equations~\eqref{SL1},~\eqref{SL2} and
\eqref{zLz}.

\subsection{Permutation group and  the star-triangle relation}
Let us consider the operators we have constructed in more details.
As was already noted, a character $\balpha$ can be written in the
form~\eqref{au} so that the representation $T^\balpha$ is
completely determined by the numbers $u_1,\ldots,u_N$ (and $\bar
u_1,\ldots,\bar u_N$ which are always implied). Respectively, the
tensor product $T^\balpha\otimes T^\bbeta$ is determined by the
numbers $u_1,\ldots,u_N$ and $v_1,\ldots,v_N$ where the latter
refer to the representation $T^\bbeta$. Let us join these
parameters into the string
\begin{gather*}
\bv=(v_1,\ldots,v_N,u_1,\ldots,u_N)\equiv
(v_1,\ldots,v_N,v_{N+1},\ldots,v_{2N})
\end{gather*}
(notice the inverse order of the parameters $\{u\}$ and $\{v\}$)
and accept the notation $\mathbb{T}^\bv$ for the tensor product
$T^\balpha\otimes T^\bbeta$. The string $\bv$
 f\/ixes not only the characters $\balpha$ and $\bbeta$
 but also the spectral parameters, $u$ and $v$  of the
Lax operators $L_1$ and $L_2$.

For a def\/initeness we will assume that the representations
$T^\balpha$ and $T^\bbeta$ are unitary. We recall that this
results in the restriction~\eqref{unitarity} which can be
represented as
\begin{gather*}
u_{k,k+1}^*+\bar u_{k,k+1}=0,\qquad v_{k,k+1}^*+\bar
v_{k,k+1}=0,\qquad k=1,\ldots,N-1.
\end{gather*}
 We will assume also that
$(v_N-u_1)^*+(\bar v_N-u_1)=v_{N,N+1}^*-\bar v_{N,N+1}=0$, so that
$v_{k,k+1}^*+\bar v_{k,k+1}=0$ for $k=1,\ldots, 2N-1$. Under this
condition the representation $\mathbb{T}^{\bv'}$ where $\bv'$ is
an arbitrary permutation of the string $\bv$ is  unitary. The
vector space of the representation $\mathbb{T}^{\bv}$ will be
denoted as $\VV_{\bv}=L^2(Z\times Z)$.

In  Section~\ref{twin} we have constructed the intertwining
operators $U_k$, equation~\eqref{U-twin}. Since one has now two
sets of such operators and also the exchange operator $S$, it is
convenient to introduce the notation
\begin{gather}\label{BUk}
\mathbb{U}_k=
\begin{cases} \II \otimes U_k, & k=1,\ldots,N-1,\\
S, & k=N,\\
U_{k-N} \otimes \II, & k=N+1,\ldots, 2N-1.
\end{cases}
\end{gather}
The operators $\mathbb{U}_k$ depend on the ``quantum numbers'' of
the space they act on, namely $ \mathbb{U}_k=
\mathbb{U}_k(v_{k+1,k})$, $v_{k+1,k}=v_{k+1}-v_k$, and
\begin{gather*}
\mathbb{U}_k: \, \VV_{\bv}\mapsto \VV_{\bv_k}, \qquad
\text{where}\qquad \bv_k=P_{k} \bv.
\end{gather*}
It will  always be  implied that the argument of the operator
$\mathbb{U}_k$ is determined by the representation
$\mathbb{T}^{\bv}$ it acts on. It can be formulated in the
following way: Let us consider the lattice $\mathcal{L}_\bv$ in
$\mathbb{C}^{2N}$ formed by vectors $\{\bv_i\}$ obtained from the
vector $\bv$  by all possible permutation of its components. For
each point $\bv_i$ we identify  the corresponding space
$\VV_{\bv_i}$. The operators $\mathbb{U}_k$ map the spaces
attached to the lattice points related by elementary permutations
to one  another. So one can omit the index $k$ of the operator but
show  the lattice points $\bv_i$ and $\bv_{i'}$ as the argument of
the operator $\mathbb{U}_k(v_{k+1,k})\to
\mathbb{U}(\bv_{i'},\bv_{i})$.
  So far we consider
operators which connect the points related by elementary
permutation. But it is obvious that one can construct the operator
which maps $\VV_{\bv_i}$ to $\VV_{\bv_j}$ where $\bv_i$ and
$\bv_j$ are two arbitrary points of the lattice. We will show that
such operator depends only on the points $\bv_i$ and $\bv_j$, and
does not depend on the path connecting these points. To this end
one has to show that the operators $\mathbb{U}_k$ satisfy the same
commutation relations as the operators of the elementary
permutation $P_{k}$: $P_{k}^2=\II$, $[P_{k},P_{n}]=0$ for
$|n-k|>1$ and $P_{k+1}P_{k}P_{k+1}=P_{k}P_{k+1}P_{k}$. Thus we
have to show that
\begin{subequations}\label{3U}
\begin{alignat}{3}
\label{U-1}
& \mathbb{U}_k \mathbb{U}_k =\II,&&&\\
\label{U-2}
& \mathbb{U}_k \mathbb{U}_n =\mathbb{U}_n \mathbb{U}_k,&&|n-k|>1 &\\
\label{U-3} & \mathbb{U}_k \mathbb{U}_{k+1}\mathbb{U}_k
=\mathbb{U}_{k+1}\mathbb{U}_k\mathbb{U}_{k+1}.&&&
\end{alignat}
\end{subequations}
Taking into account equations~\eqref{UUU} which hold for the
operators $U_k$ one easily f\/igures out that it is suf\/f\/icient
to check only those of  equations~\eqref{3U} which involve the
operator $\mathbb{U}_N$.  The f\/irst equation,
$\mathbb{U}_N\mathbb{U}_N=\II$, is obvious. The second one,
$\mathbb{U}_N \mathbb{U}_k=\mathbb{U}_k \mathbb{U}_N$, $|N-k|>1$,
follows immediately from equations~\eqref{U-twin} and \eqref{defS}
if one takes into account that $D_{k+1,k}^{z} (w^{-1}
z)_{N1}=D_{i+1,i}^{w} (w^{-1} z)_{N1}=0$ for $k>1$ and $i<N-1$.

 The last
identity, equation~\eqref{U-3}, is nothing else as a slightly
camouf\/laged version of the integral identity known as the
star-triangle relation\footnote{The star-triangle relation is, in
some sense, a key feature of the Yang--Baxter equation~\eqref{YB},
see \cite{PY} for a~nice review.}. If we restore  the dependence
on the spectral parameters it takes the form
\begin{subequations}\label{st-tr}
\begin{gather}
\label{sta} \mathbb{U}_N(b) \mathbb{U}_{N+1}(a+b) \mathbb{U}_N(a)
=
\mathbb{U}_{N+1}(a)\mathbb{U}_{N}(a+b)\mathbb{U}_{N+1}(b), \\
\label{stb} \mathbb{U}_N(b) \mathbb{U}_{N-1}(a+b) \mathbb{U}_N(a)
= \mathbb{U}_{N-1}(a)\mathbb{U}_{N}(a+b)\mathbb{U}_{N-1}(b).
\end{gather}
\end{subequations}

Let us consider equation~\eqref{sta}. We recall  that
$\mathbb{U}_{N+1}(a)= (D_{21}^{z})^{a} (\bar D_{21}^{\bar
z})^{\bar a}$. After the  change of variables~\eqref{ztox} this
operator takes the form
$\mathbb{U}_{N+1}(a)={\partial}^a\,\bar\partial^{\bar a}$, where
$\partial\equiv\partial_x$ and $x=x_{21}$. It  turn, the matrix
element $(w^{-1}z)_{N1}$ is  a linear function of  $x$,
\begin{gather*}
(w^{-1}z)_{N1}=Cx+D=C(x+x_0),\qquad x_0=D/C,
\end{gather*}
where  $C$ and $D$ depend on other variables and can be treated as
constants. It is easy to see that equation~\eqref{sta} is
equivalent to the statement about commutativity of  the operators
\begin{gather*}
G_a=(x^a\,\partial^a) (\bar x^{\bar a} \bar\partial^{\bar a})
\qquad \text{and} \qquad H_b=(\partial^b x^b) (\bar\partial^{\bar
b} \bar x^{\bar b}).
\end{gather*}
The latter follows from the observation that both operators are
some functions  of the operators $x\partial$ and $\bar
x\bar\partial$, $G_a=g_a(x\partial,\bar x\bar\partial)$ and
$H_b=h_a(x\partial,\bar x\bar\partial)$. This line of reasoning is
due to Isaev~\cite{Isaev}.

The property of  commutativity of the operators $H_b$ and $G_a$
can be expressed as the integral identity which presents a more
conventional form   of the star-triangle relation
\begin{gather}
A(a)A(b)\int d^2\xi \frac{1}{[x-\xi]^{a+1}[\xi-\zeta]^{-a-b}[\xi-x']^{b+1}}\nonumber\\
\qquad{}=
\frac{A(a+b)}{[x-\zeta]^{-b}[x-x']^{1+a+b}[x'-\zeta]^{-a}}.\label{star-triangle}
\end{gather}
Here the function $A(a)$ is def\/ined in equation~\eqref{prop} and
$[x]^a= x^{a} \bar x^{\bar a}$. Using
equation~\eqref{star-triangle} and the integral
representation~\eqref{prop} for the operators $U_k$ it is
straightforward to verify equations~\eqref{st-tr}.

Thus we have proved that the operators $\mathbb{U}_k$~\eqref{BUk}
satisfy the commutation relations~\eqref{3U}. Now one can
construct the operator $\mathbb{U}(\bv_j,\bv_i)$ which maps
$\VV_{\bv_i}\mapsto \VV_{\bv_j}$ where vectors~$\bv_i$,~$\bv_j$
belong to the lattice $\mathcal{L}_\bv$. By def\/inition, the
vectors  $\bv_i$, $\bv_j$ are related by some permutation~$P(ij)$,
$\bv_j=P(ij)\bv_i$. The permutation $P(ij)$ can be always
represented as the product of the elementary permutations,
$P(ij)=P_{i_m}\cdots P_{i_1}$. Then we def\/ine the operator
$\mathbb{U}(\bv_j,\bv_i)$ as
\begin{gather*}
\mathbb{U}(\bv_j,\bv_i)=\mathbb{U}_{i_m}\cdots \mathbb{U}_{i_1}.
\end{gather*}
Due to equations~\eqref{3U} the operator $\mathbb{U}(\bv_j,\bv_i)$
does not depend on the way the decomposition of the permutation
$P(ij)$ onto elementary permutations is done. The operator
$\mathbb{U}(\bv',\bv)$ intertwines the representations
$\mathbb{T}^{\bv}$ and $\mathbb{T}^{\bv'}$
\begin{gather*}
\mathbb{U}(\bv',\bv)\mathbb{T}^{\bv}=\mathbb{T}^{\bv'}\mathbb{U}(\bv',\bv)
\end{gather*}
and interchanges the parameters in the product of Lax operators as
follows
\begin{gather*}
\mathbb{U}(\bv',\bv)L_1(v_{N+1},\ldots,v_{2N})L_2(v_{1},\ldots,v_N)=
L_1(v_{N+1}^\prime,\ldots,v_{2N}^\prime)
L_2(v_{1}^\prime,\ldots,v_N^\prime) \mathbb{U}(\bv',\bv).
\end{gather*}
It follows from equation~\eqref{RLL-1} that the operator
$\mathbb{R}_{12}(u-v)$ can be identif\/ied with
$\mathbb{U}(\bv',\bv)$ for the special $\bv'$. Namely, one gets
\begin{gather}\label{RU}
\mathbb{R}_{12}(u-v)=\mathbb{U}(\bv',\bv),
\end{gather}
where
\begin{gather*}
\bv=(v_1,\ldots,v_N,u_1,\ldots,u_N),\qquad
\bv'=(u_1,\ldots,u_N,v_1,\ldots,v_N).
\end{gather*}
As a consequence the $\CR$-operator takes the form
\begin{gather*}
\CR_{12}(u-v)= P_{12} \mathbb{U}(\bv',\bv).
\end{gather*}
Note, that since each operator $\mathbb{U}_k$ depends on the
dif\/ference $v_{k,k+1}$, the operator $\mathbb{U}(\bv',\bv)$
depends on the dif\/ferences $\sigma_{k,k+1}$ and $\rho_{k,k+1}$,
which specify the representations $T^\balpha$ and $T^\bbeta$ in
the tensor product $T^\balpha\otimes T^\bbeta$,
 and the
spectral parameter $u-v$.

It is quite easy to show that the constructed $\CR$-operator
satisf\/ies the Yang--Baxter relation~\eqref{YB}. The latter can
be rewritten in the form
\begin{gather}
\left(P_{23}P_{12}P_{23}\right) \mathbb{R}_{23}(\vec{u},\vec{v})
\mathbb{R}_{12}(\vec{u},\vec{w})
\mathbb{R}_{23}(\vec{v},\vec{w})\nonumber\\
\qquad{}=\left(P_{12}P_{23}P_{12}\right)
\mathbb{R}_{12}(\vec{v},\vec{w}) \mathbb{R}_{23}(\vec{u},\vec{w})
\mathbb{R}_{12}(\vec{u},\vec{v}),\label{PYB}
\end{gather}
where we  have  shown all arguments of $\mathbb{R}$-operators
explicitly, that is
$\mathbb{R}_{12}(u-v)=\mathbb{R}_{12}(\vec{u},\vec{v})$,
$\vec{u}=(u_1,\ldots,u_N)$ and so on.
 Since
$P_{23}P_{12}P_{23}=P_{12}P_{23}P_{12}$ one has to check that the
product of $\mathbb{R}$-operators in the l.h.s and r.h.s are
equal. One can easily f\/ind that the operators on the both sides
result in the same permutation of the parameters
$(\vec{u},\vec{v},\vec{w})\to(\vec{w},\vec{v},\vec{u}) $. Since
the operators are constructed from the operators $\mathbb{U}_k$
which obey the relations~\eqref{3U}, these operators, as was
explained earlier, are equal.

\section[Factorizing operators and Baxter $\mathcal{Q}$-operators]{Factorizing operators
and Baxter $\boldsymbol{\mathcal{Q}}$-operators}\label{sect:fq}

In the previous section we have obtained the expression for the
operator $\mathbb{R}_{12}$,~equation~\eqref{RU}. For the
construction of  the Baxter $\mathcal{Q}$-operators it is quite
useful to represent the  $\mathbb{R}$-operator in the factorized
form~\eqref{RRk}. Each operator $\mathbb{R}^{(k)}$ interchanges
the components $u_k$ and $v_k$ of the Lax operators,
equation~\eqref{RkL} and can be expressed in terms of the
elementary permutation opera\-tors~$\mathbb{U}_k$ as follows
\begin{gather}\label{ERk}
\mathbb{R}^{(k)}=(\mathbb{U}_{N+k-1}\cdots
\mathbb{U}_{N+1})(\mathbb{U}_{k}\cdots \mathbb{U}_{{N-1}})
\mathbb{U}_N (\mathbb{U}_{N-1}\cdots
\mathbb{U}_{k})\,(\mathbb{U}_{N+1}\cdots \mathbb{U}_{N+k-1})
\end{gather}
Taking into account the def\/inition of the operators
$\mathbb{U}_k$, equation~\eqref{BUk}, it is straightforward to
check that the operator $\mathbb{R}^{(k)}$ satisf\/ies
equation~\eqref{RkL}. Indeed, sequence of the operators
in~\eqref{ERk} results in the following permutation of the
parameters $\vec{u}$ and $\vec{v}$ in the product of Lax operators{\samepage
\begin{gather*}
(\cdots v_k,v_{k+1}\cdots v_N|u_1\cdots u_{k-1},
u_k\cdots)\xrightarrow{\mathbb{U}_{N+1}\cdots
\mathbb{U}_{N+k-1}}(\cdots v_k,v_{k+1}\cdots v_N|u_k,u_1\cdots
u_{k-1}\cdots)
\\
\xrightarrow{\mathbb{U}_{N-1}\cdots \mathbb{U}_{k} } (\cdots
v_{k+1}\cdots v_N, v_k|u_k,u_1\cdots
u_{k-1}\cdots)\xrightarrow{\mathbb{U}_N}
(\cdots v_{k+1}\cdots v_N, u_k|v_k,u_1\cdots u_{k-1}\cdots)\\
\xrightarrow{\mathbb{U}_{k}\cdots \mathbb{U}_{N-1}}
(\cdots u_k,v_{k+1}\cdots v_N|v_k,u_1\cdots u_{k-1}\cdots)\\
\xrightarrow{\mathbb{U}_{N+k-1}\cdots \mathbb{U}_{N+1}} (\cdots
u_k,v_{k+1}\cdots v_N|u_1\cdots u_{k-1},v_k\cdots),
\end{gather*}
where we have displayed  the relevant arguments only.}

It is easy to show that the operator~\eqref{ERk}
 intertwines the representations $T^\balpha\otimes T^\bbeta$ and
$T^{\balpha_{k,\lambda}}\otimes T^{\bbeta_{k,-\lambda}}$, where
\begin{gather}\label{char}
\balpha_{k,\lambda}(h)=h_{kk}^{-\lambda}\,\bar
h_{kk}^{-\bar\lambda}\,\balpha(h),\qquad
\bbeta_{k,-\lambda}(h)=h_{kk}^{\lambda}\,\bar
h_{kk}^{\bar\lambda}\,\bbeta(h),
\end{gather}
with $\lambda=u_k-v_k$, $\bar\lambda=\bar u_k-\bar v_k$. The
operator $\mathbb{R}^{(k)}$ is completely determined by the
charac\-ters~$\balpha$,~$\bbeta$ and the parameter
$\lambda,(\bar\lambda)$, which we will refer to as the spectral
parameter, i.e.\  $\mathbb{R}^{(k)}=
\mathbb{R}^{(k)}(\lambda|\balpha,\bbeta)$. Henceforth  we accept
the shorthand notation,
$\mathbb{R}^{(k)}(\lambda|\balpha,\bbeta)\to\mathbb{R}^{(k)}(\lambda)$,
omitting the  dependence on the characters $\balpha$, $\bbeta$. To
avoid misunderstanding we stress that the product of operators
$\mathbb{R}^{(i)}(\mu)\mathbb{R}^{(k)}(\lambda)$ reads in explicit
form  as
\begin{gather}\label{remark}
\mathbb{R}^{(i)}(\mu)\mathbb{R}^{(k)}(\lambda)\equiv
\mathbb{R}^{(i)}(\mu|\balpha_{k,\lambda},\bbeta_{k,-\lambda})\mathbb{R}^{(k)}(\lambda|\balpha,\bbeta).
\end{gather}
In a full analogy with $\CR$-operator we accept the notation
$\mathbb{R}^{(k)}_{ab}$ for the operator which acts  nontrivially
on the tensor product of spaces $\VV_a$ and $\VV_b$. The
expression~\eqref{RRk}  for the $\CR$-operator can be written in
the form
\begin{gather}\label{R-factform}
\CR_{12}(u-v)=P_{12}\mathbb{R}_{12}(u-v)=P_{12}\mathbb{R}^{(1)}_{12}(u_1-v_1)
\mathbb{R}^{(2)}_{12}(u_2-v_2)
\cdots
\mathbb{R}^{(N)}_{12}(u_N-v_N).
\end{gather}
We recall that $u_k=u-\sigma_k$ and $v_k=v-\rho_k$, where the
parameters $\sigma$ and $\rho$ def\/ine the characters~$\balpha$
and $\bbeta$, see equations~\eqref{dh} and \eqref{rs}.

The operators $\mathbb{R}^{(k)}_{ab}(\lambda)$ possess a number of
remarkable properties
\begin{subequations}\label{propR}
\begin{gather}
\label{unity} \mathbb{R}^{(k)}_{12}(0) = \II,
\\
\label{p1}
\mathbb{R}^{(k)}_{12}(\lambda)\mathbb{R}^{(k)}_{12}(\mu) =
\mathbb{R}^{(k)}_{12}(\lambda+\mu),
\\
\label{R-R}
\mathbb{R}_{12}^{(k)}(\lambda)\mathbb{R}_{23}^{(j)}(\mu) =
\mathbb{R}_{23}^{(j)}(\mu)\mathbb{R}_{12}^{(k)}(\lambda), \qquad
j\neq k,\\
\label{R-R-R}
\mathbb{R}_{12}^{(k)}(\lambda)\mathbb{R}_{23}^{(k)}(\lambda+\mu)\mathbb{R}_{12}^{(k)}(\mu)
 =
\mathbb{R}_{23}^{(k)}(\mu)\mathbb{R}_{12}^{(k)}(\lambda+\mu)\mathbb{R}_{23}^{(k)}(\lambda),\\
\label{R-2-R}
\mathbb{R}^{(k)}_{12}(\lambda-\sigma_k+\rho_k)\mathbb{R}^{(i)}_{12}(\lambda-\sigma_i+\rho_i)
=
\mathbb{R}^{(i)}_{12}(\lambda-\sigma_i+\rho_i)\mathbb{R}^{(k)}_{12}(\lambda-\sigma_k+\rho_k).
\end{gather}
\end{subequations}
Equations~\eqref{unity} and \eqref{p1} follow from
equations~\eqref{ERk} and \eqref{defS}, while to prove the last
three equations it is suf\/f\/icient to check that the operators
on both sides result in the same permutation of the parameters
$\vec{u}$, $\vec{v}$, $\vec{w}$, (see equation~\eqref{PYB}).
Namely, the equations~\eqref{R-R}--\eqref{R-2-R}
 are the deciphered form of  the equations
\begin{gather*}
\mathbb{R}_{12}^{(k)}\mathbb{R}_{23}^{(j)} =
\mathbb{R}_{23}^{(j)}\mathbb{R}_{12}^{(k)}, \qquad
\mathbb{R}_{12}^{(k)}\mathbb{R}_{23}^{(k)}\mathbb{R}_{12}^{(k)}
 =
\mathbb{R}_{23}^{(k)}\mathbb{R}_{12}^{(k)}\mathbb{R}_{23}^{(k)}
\qquad \text{and} \qquad
\mathbb{R}_{12}^{(k)}\mathbb{R}_{12}^{(i)}=\mathbb{R}_{12}^{(i)}\mathbb{R}_{12}^{(k)}.
\end{gather*}
Let us notice that equation~\eqref{R-2-R} indicates that the
operators $\mathbb{R}^{k}(u_k-v_k)$ in the expression for the
$\CR$-matrix,~equation~\eqref{R-factform}, can stand in arbitrary
order.

We remark here that YBE for the $\CR$-operator~\eqref{RRk} is the
corollary of the properties~\eqref{R-R}, \eqref{R-R-R} and
\eqref{R-2-R} of the operators $\mathbb{R}^{(k)}$. One can expect
that these equations will hold for the $\CR$ matrix for the
generic representation of $sl(N)$ algebra since they express the
property of consistency  of equations~\eqref{RkL}. At the same
time the possibility to represent the operators~$\mathbb{R}^{(k)}$
as the product of  elementary operators $\mathbb{U}_k$ is a
specif\/ic feature of the principal series representations. Taking
in mind this possibility we give here the alternative  proof of
YBE. We recall that  YBE is equivalent to the following identity
for the operators $\mathbb{R}_{ik}$ (see discussion after
equation~\eqref{PYB})
\begin{gather}\label{RRR}
\mathbb{R}_{23}(u-v)\mathbb{R}_{12}(u-w)\mathbb{R}_{23}(v-w)=
\mathbb{R}_{12}(v-w)\mathbb{R}_{23}(u-w)\mathbb{R}_{12}(u-v).
\end{gather}
Using the factorized form~\eqref{R-factform} for the operator
$\mathbb{R}_{ik}$ and using equations~\eqref{R-R} and
\eqref{R-2-R} one can bring the  l.h.s and r.h.s of
equation~\eqref{RRR} into the form
\begin{gather*}
\mathbb{R}^{(1)}_{23}(u_1-v_1)\mathbb{R}^{(1)}_{12}(u_1-w_1)\mathbb{R}^{(1)}_{23}(v_1-w_1)
\cdots
\mathbb{R}^{(N)}_{23}(u_N-v_N)\\
\qquad{}\times
\mathbb{R}^{(N)}_{12}(u_N-w_N)\mathbb{R}^{(N)}_{23}(v_N-w_N)
\end{gather*}
and{\samepage
\begin{gather*}
\mathbb{R}^{(1)}_{12}(v_1-w_1)\mathbb{R}^{(1)}_{23}(u_1-w_1)\mathbb{R}^{(1)}_{12}(u_1-v_1)
\cdots
\mathbb{R}^{(N)}_{12}(v_N-w_N)\\
\qquad{}\times
\mathbb{R}^{(N)}_{23}(u_N-w_N)\mathbb{R}^{(N)}_{12}(u_N-v_N),
\end{gather*}
respectively. By virtue of equation~\eqref{R-R-R} one f\/inds that
they are equal.}

So we have shown that if the operators $\mathbb{R}^{(k)}$ obey
equations~\eqref{R-R}--\eqref{R-2-R} then the
$\CR$-matrix~\eqref{R-factform} satisf\/ies YBE~\eqref{YB}.
Moreover, the operators $\mathbb{R}^{(k)}$ can be considered as  a
special case of the $\CR$-operator. Namely, let us take the
character $\bbeta(h)$ to be $\bbeta(h)=\balpha_{k,\lambda}(h)$.
Then it follows from equations~\eqref{char} that the operator
\begin{gather}\label{defR}
\CR_{12}^{(k)}(\lambda)= P_{12}\mathbb{R}_{12}^{(k)}(\lambda)
\end{gather}
 is a
$SL(N,\mathbb{C})$ invariant operator,
\begin{gather*}
\CR_{12}^{(k)}(\lambda) \left(T^\balpha\otimes
T^{\balpha_{k,\lambda}}\right)(g)= \left(T^\balpha\otimes
T^{\balpha_{k,\lambda}}\right)(g) \CR_{12}^{(k)}(\lambda).
\end{gather*}
Let us note
 that the operators $\mathbb{R}^{(k)}(\lambda)$ depend only on the part of parameters $\sigma_i$
and $\rho_i$ which specify the characters $\balpha$ and $\bbeta$.
For instance, an inspection of equation~\eqref{ERk} shows that the
operator $\mathbb{R}^{(N)}(\lambda) $ depends only on the
parameters $\sigma_{k,k+1}=\sigma_k-\sigma_{k+1}$ (the character
$\balpha$),
($\mathbb{R}^{(N)}(\lambda)\equiv\mathbb{R}^{(k)}(\lambda|\sigma_{12},\ldots,\sigma_{N-1,N})$),
while the operator $\mathbb{R}^{(1)}(\lambda) $ depends on
$\rho_{k,k+1}$ (the character $\bbeta$), and similarly for others.
It means that the condition $\bbeta(h)=\balpha_{k,\lambda}(h)$
does not reduce the number of the independent parameters the
operator $\mathbb{R}^{(k)}$  depends on.

Further, let us return to equation~\eqref{R-factform} and put
$u_k-v_k=\lambda$. Then, provided that
$\bbeta(h)=\alpha_{k,\lambda}(h)$ one f\/inds $u_i-v_i=0$ unless
$i=k$. Taking into account equation~\eqref{unity} and $u-v=\sum_k
(u_k-v_k)/N=\lambda/N$ one derives
\begin{gather*}
\CR_{12}^{(k)}(\lambda)=P_{12}\mathbb{R}_{12}^{(k)}(\lambda)=
\CR_{12}\left(\frac{\lambda}{N}\right)\Big|_{\bbeta(h)=\balpha_{k,\lambda}(h)}.
\end{gather*}
It means that the operator $\CR^{(k)}$ coincides  with the
$\CR$-operator at the ``special point''. In case
$\lambda+\bar\lambda^*=0$, the operator $\CR^{(k)}_{12}(\lambda)$
is  a unitary operator on the Hilbert space
$\VV_1\otimes\VV_2=L^2(Z\times Z)$ which carries the
representation $T^\balpha\otimes T^{\balpha_{k,\lambda}}$,
\begin{gather*}
\left(\CR^{(k)}_{12}(\lambda)\right)^\dagger\CR^{(k)}_{12}(\lambda)=\II.
\end{gather*}
Moreover, these operators for dif\/ferent $k$ are unitary
equivalent, for example
\begin{gather}\label{unit-equiv}
\CR^{(k)}_{12}(\lambda)=W_k^\dagger \CR^{(N)}_{12}(\lambda)  W_k .
\end{gather}
The operator $W_k$ can be easily read of\/f from
equation~\eqref{ERk}. It factorizes into the product of two
operators, $W_k=W^{(1)}_k\, {\widetilde W}^{(2)}_k$, where the
unitary operators $W^{(1)}_k$ and $ {\widetilde W}^{(2)}_k$ act in
the spaces $\VV_1$ and $\VV_2$, respectively,
\begin{gather*}
W^{(1)}_k= \mathbb{U}_{2N-1}(\sigma_{k-1,N})\cdots \mathbb{U}_{N+k-1}(\sigma_{k-1,k}),\\
{\widetilde W}^{(2)}_k=
 \mathbb{U}_{N-1}(\sigma_{k,N}+\lambda)\cdots\mathbb{U}_{k}(\sigma_{k,k+1}+\lambda).
\end{gather*}
Here we display explicitly the arguments of the intertwining
operators.

Let us consider  the homogeneous spin chain of length $L$ and
def\/ine
 the operators $\mathcal{Q}_k(\lambda)$ by\footnote{We recall our
  convention is to display only the holomorphic parameter $\lambda$, i.e.\
the operator
$\mathcal{Q}_k(\lambda)\equiv\mathcal{Q}_k(\lambda,\bar\lambda)$,
where $\lambda-\bar\lambda$ is integer.}
\begin{gather}\label{Q}
\mathcal{Q}_k(\lambda+\sigma_k)=\Tr_0\left\{
\CR_{10}^{(k)}(\lambda)\cdots\CR_{L0}^{(k)}(\lambda)\right\}.
\end{gather}
The operators $\mathcal{Q}_k$ act on the Hilbert space
$\VV_1\otimes\cdots\otimes\VV_L= L^2(Z\times\cdots\times Z)$. It
is assumed that the representation $T^\balpha$ of the group
$SL(N,\mathbb{C})$ is def\/ined at each site. The trace is taken
over the auxiliary space $\VV_0=L^2(Z)$ and exists (we give below
the explicit realization of  $\mathcal{Q}_k(\lambda)$ as an
integral operator).

The operators~\eqref{Q} can be identif\/ied as the Baxter
operators for the noncompact $SL(N,\mathbb{C})$ invariant spin
magnet. Since the operators $\CR_{i0}^{(k)}(\lambda)$ coincide
with the $\CR_{i0}$-operator  for the special choice of auxiliary
space one concludes that the Baxter operators
$\mathcal{Q}_k(\lambda)$ commute with each other for dif\/ferent
values of a spectral parameter
\begin{gather*}
[\mathcal{Q}_k(\lambda),\mathcal{Q}_i(\mu)]=0.
\end{gather*}
As follows from equation~\eqref{unit-equiv} all Baxter operators
are unitary equivalent, namely
\begin{gather}   \label{similarity}
\mathcal{Q}_k(\lambda+\sigma_k)=\left(\prod_{i=i}^L W_k^{(i)}
\right)^\dagger \mathcal{Q}_N(\lambda+\sigma_N)\left(\prod_{i=i}^L
W_k^{(i)} \right).
\end{gather}
This property is a special feature of the principal series
$SL(N,\mathbb{C})$ magnets (see also~\cite{SL2C} where the
$SL(2,\mathbb{C})$ magnet was studied in detail). For
conventional (inf\/inite-dimensional) $sl(N)$ magnets the
property~\eqref{similarity} does not hold
(see~\cite{SD-2,DM-I,DM-II}).

Another property of the Baxter operators which we want to mention
is related to the factorization property of the transfer matrix.
The latter is def\/ined as
\begin{gather}\label{t-matrix}
\mathsf{T}_\bbeta(\lambda)=\Tr_0\left\{
\CR_{10}(\lambda)\cdots\CR_{L0}(\lambda) \right\},
\end{gather}
where the trace is taken over the auxiliary space and the index
$\bbeta$ refers to the representation $T^\bbeta$ which is
def\/ined on the space $\VV_0$. We will suppose that the trace
exists (this is true at least  for $N=2$~\cite{SL2C}).  Under this
assumption one can prove that the transfer matrix factorizes in
the product of $\mathcal{Q}_k$ operators as follows
\begin{gather*}
\mathsf{T}_\bbeta(\lambda)=\mathcal{Q}_1(\lambda+\rho_1)\mathcal{P}^{-1}
\mathcal{Q}_2(\lambda+\rho_2)\mathcal{P}^{-1}\cdots
\mathcal{P}^{-1} \mathcal{Q}_N(\lambda+\rho_N),
\end{gather*}
where $\mathcal{P}$ is the operator of cyclic permutation
\begin{gather*}
 \mathcal{P}\Phi(z_1,\ldots,z_L)=\Phi(z_L,z_1,\ldots,z_{L-1})
\end{gather*}
and the parameters $\rho_k$ determine the character $\bbeta$,
equations~\eqref{dh}, \eqref{rs}. The proof repeats that given
in~\cite{DM-II} and is based on disentangling of the
trace~equation~\eqref{t-matrix} with the help of the
relation~\eqref{R-R} for  $\mathbb{R}^{(k)}$ operators. The
analysis of analytical properties of the Baxter
$\mathcal{Q}$-operators and a derivation of the
$\mathsf{T}$-$\mathcal{Q}$ relation will be given elsewhere.

\subsection{Integral representation}

In this subsection we give the integral representation  for the
operators in question. Due to equations~\eqref{unit-equiv} and
\eqref{similarity} it suf\/f\/ices to consider the operators
$\mathcal{Q}_N$ and $\mathbb{R}^{(N)}$ only. The latter can be
represented as the product of three operators,
$\mathbb{R}^{(N)}(\lambda)=\widetilde{\mathcal{X}_1}(\lambda)
S(\lambda) \mathcal{X}_1$. The operator~$S(\lambda)$  def\/ined in
equation~\eqref{defS} is   an multiplication operator
\begin{gather}\label{ss}
[S(\lambda)\Phi](z,w)=[w,z]^\lambda\,\Phi(z,w),
\end{gather}
where
$[w,z]^\lambda\equiv(w^{-1}z)_{N1}^{\lambda}((w^{-1}z)_{N1}^*)^{\bar
\lambda}$. The operator $\mathcal{X}_1$ is given by a product of
the operators $(U_1\otimes\II)(U_2\otimes\II)\cdots(U_{N-1}\otimes
\II)$, see equations~\eqref{ERk} and \eqref{BUk}. Making use of
equation~\eqref{prop} one gets
\begin{gather}\label{x1}
[\mathcal{X}_1\Phi](z,w)= \prod_{k=1}^{N-1}\,A(\sigma_{kN}) \int
d^{2}\zeta_k\, {[z_{k+1,k}-\zeta_k]^{-1-\sigma_{kN}}}
\Phi(z_{\boldsymbol{\zeta}},w).
\end{gather}
The factor $A$ is def\/ined  in equation~\eqref{A-factor},
$\sigma_{kN}=\sigma_k-\sigma_N$,
\begin{gather*}
z_{\boldsymbol{\zeta}}=
z\big(1+(\zeta_1-z_{21})e_{21}\big)\cdots\big(1+(\zeta_{N-1}-z_{NN-1})e_{NN-1}\big)\\
\phantom{z_{\boldsymbol{\zeta}}}{}=z\left(1+\sum_{k=1}^{N-1}
(\zeta_k-z_{k+1,k})e_{k+1,k}\right).
\end{gather*}
and we recall that $[z]^a\equiv z^{a} \bar z^{\bar a}$. Similarly,
for the operator $\widetilde{\mathcal{X}_1}(\lambda)$ one derives
\begin{gather}\label{tx1}
[\widetilde{\mathcal{X}_1}(\lambda)\Phi](z,w)=
\prod_{k=1}^{N-1}\,A(\sigma_{Nk}+\lambda) \int d^{2}\zeta_k\,
{[z_{k+1,k}-\zeta_k]^{-(1+\lambda+\sigma_{Nk})}}\Phi(\tilde{z}_{\boldsymbol{\zeta}},w),
\end{gather}
with
\begin{gather*}
\tilde{z}_{\boldsymbol{\zeta}}=
z\left(1+(\zeta_{N-1}-z_{NN-1})e_{NN-1}\right)\cdots\left(1+(\zeta_{1}-z_{21})e_{21}\right).
\end{gather*}
Thus  the operator $\mathbb{R}^{(N)}(\lambda)$ can be represented
in the following form
\begin{gather}
[\mathbb{R}_{12}^{(N)}(\lambda)\Phi](z,w) =\prod_{k=1}^{N-1}
A(\sigma_{kN})A(\sigma_{Nk}+\lambda) \int d^{2}\zeta_k\int
d^{2}\zeta'_k\nonumber\\
\phantom{[\mathbb{R}_{12}^{(N)}(\lambda)\Phi](z,w)=}{}\times[z_{k+1,k}-\zeta_k]^{-(1+\lambda+\sigma_{Nk})}[\zeta_k-\zeta'_k]^{-(1+\sigma_{kN})}
[w,\tilde z_\zeta]^{\lambda} \Phi(z_{\zeta\zeta'},w),\label{RN}
\end{gather}
where
\begin{gather*}
z_{\zeta\zeta'}=\tilde{z}_{\boldsymbol{\zeta}}
\left(1+(\zeta'_1-\zeta_1)e_{21}\right)\cdots\left(1+(\zeta'_{N-1}-\zeta_{N-1})e_{NN-1}\right).
\end{gather*}

To obtain the integral representation for the Baxter operator
$\mathcal{Q}_N$ we notice that the operator
$\CR^{(N)}_{12}(\lambda)$ leaves the second argument of the
function $\Phi(z,w)$ intact. Therefore the integral kernel of the
operator $\CR^{(N)}_{12}(\lambda)$,
\begin{gather*}
[\mathbb{R}_{12}^{(N)}(\lambda)\Phi](z,w)=\int Dz' Dw'
R_\lambda(z,w|z',w')\Phi(z',w'),
\end{gather*}
where $Dz'=\prod\limits_{k>i} d^2z'_{ki}$,
$Dw'=\prod\limits_{k>i}\,d^2w'_{ki}$, has the following form
\begin{gather*}
R_\lambda(z,w|z',w')=R_\lambda(z,w|z')\delta(w-w'),
\end{gather*}
with $\delta(w-w')=\prod\limits_{k>i} \delta^2(w_{ki}-w'_{ki})$.
Making use of equations~\eqref{defR} and \eqref{Q} one f\/inds
that the kernel $Q_\lambda^{(N)}$ of the operator $\mathcal{Q}_N$
\begin{gather*}
[\mathcal{Q}_N(\lambda+\sigma_N)\Phi](z_1,\ldots,z_L)=\int
\prod_{k=1}^L Dz'_k\,
Q_\lambda^{(N)}(z_1,\ldots,z_L|z'_1,\ldots,z'_L)\,\Phi(z'_1,\ldots,z'_L)\,.
\end{gather*}
for a spin chain of a length  $L$ has the following form
\begin{gather*}
Q_\lambda^{(N)}(z_1,\ldots,z_L|z'_1,\ldots,z'_L)=\prod_{k=1}^L
R_\lambda(z_k,z_{k+1}|z'_{k+1}),
\end{gather*}
where the periodic boundary conditions are implied $z_{L+1}=z_1$.
Further, taking into account equations~\eqref{ss}, \eqref{x1},
\eqref{tx1} one can represent the operator $\mathcal{Q}^{(N)}$ in
the form
\begin{gather}  \label{QX}
\mathcal{Q}^{(N)}(\lambda+\sigma_N)=\widebar{\mathcal{Q}}(\lambda+\sigma_N)
\mathcal{X}_1\mathcal{X}_2\cdots\mathcal{X}_L.
\end{gather}
The operator $\mathcal{X}_k$, which acts non-trivially only on the
$k$-th component in the tensor product
$\VV_1\otimes\VV_2\otimes\cdots\otimes \VV_L$, is given by
equation~\eqref{x1}. In turn, for the operator
$\widebar{\mathcal{Q}}(\lambda)$ one obtains
\begin{gather*}
[\widebar{\mathcal{Q}}(\lambda)\Phi](z_1,\ldots,z_L)=q^L(\lambda)\prod_{k=1}^L
\left(\prod_{j=1}^{N-1} \int d^2
\zeta_{k,j}{[(z_k)_{j+1,j}-\zeta_{k,j}]^{-(1+\lambda-\sigma_{j})}}\right)
\nonumber\\
\phantom{[\widebar{\mathcal{Q}}(\lambda)\Phi](z_1,\ldots,z_L)=}{}
\times[z_{k+1},(\tilde z_{k})_{\zeta_k} ]^{\lambda-\sigma_N}\,
\Phi((\tilde z_{L})_{\zeta_L},(\tilde
z_{1})_{\zeta_1},\ldots,(\tilde z_{L-1})_{\zeta_{L-1}}),
\end{gather*}
where
\begin{gather*}
q(\lambda)=\prod_{j=1}^{N-1}A(\lambda-\sigma_j).
\end{gather*}

\subsection[$SL(2,\mathbb{C})$ magnet]{$\boldsymbol{SL(2,\mathbb{C})}$ magnet}

Closing this section we give the explicit expression for the
operator $\mathbb{R}^{(N)}$ and the Baxter $\mathcal{Q}$-operator
for the $SL(2,\mathbb{C})$ spin magnet. For $N=2$
equation~\eqref{RN} takes the form
\begin{gather*}
\big[\mathbb{R}_{12}^{(N=2)}(\lambda) \Phi\big](z,w) =
A(\sigma_{12})A(\lambda-\sigma_{12}) \int d^{2}\zeta\int
d^{2}\zeta'\nonumber\\
\phantom{\big[\mathbb{R}_{12}^{(N=2)}(\lambda) \Phi\big](z,w)=}{}
\times[z-\zeta]^{-(1+\lambda-\sigma_{12})}[\zeta-\zeta']^{-(1+\sigma_{12})}
[\zeta-w]^{\lambda} \Phi(\zeta',w),
\end{gather*}
where we put $z_{12}=z$, $w_{12}=w$. Integrating over $\zeta$ with
help of equation~\eqref{star-triangle} one derives
\begin{gather*}
\big[\mathbb{R}_{12}^{(N=2)}(\lambda) \Phi\big](z,w)
={A(\lambda)}{[z-w]^{\sigma_{12}}}\int d^{2}\zeta'
[z-\zeta']^{-1-\lambda}
[\zeta'-w]^{\lambda-\sigma_{12}}\, \Phi(\zeta',w)\nonumber\\
\phantom{\big[\mathbb{R}_{12}^{(N=2)}(\lambda) \Phi\big](z,w)}{} =
A(\lambda)\int d^2\alpha
[\alpha]^{-1-\lambda}\,[1-\alpha]^{\lambda-\sigma_{12}}
\Phi(z(1-\alpha)+w\alpha,w).
\end{gather*}
Similarly, for the operator $\mathbb{R}^{(1)}_{12}$ one f\/inds
\begin{gather*}
\big[\mathbb{R}_{12}^{(1)}(\lambda)\Phi\big](z,w)
={A(\lambda)}{[z-w]^{\rho_{12}}}\int d^{2}\zeta'\,
[w-\zeta']^{-1-\lambda}
[z-\zeta']^{\lambda-\rho_{12}} \Phi(z,\zeta')\nonumber\\
\phantom{\big[\mathbb{R}_{12}^{(1)}(\lambda)\Phi\big](z,w)}{}
=(-1)^{\lambda-\bar\lambda} {A(\lambda)}\int d^2\alpha
[\alpha]^{-1-\lambda}[1-\alpha]^{\lambda-\rho_{12}}
\Phi(z,w(1-\alpha)+z\alpha).
\end{gather*}
With the help of equations~\eqref{R-factform} and \eqref{remark}
one can easily f\/ind that the integral  kernel  of the operator
$\CR_{12}(\lambda)$
\begin{gather*}
\left[\CR_{12}(\lambda)\Phi\right](z,w)=\int d^2z'
d^2w'\,R_\lambda(z,w|z',w')\Phi(z',w')
\end{gather*}
takes the form
\begin{gather*}
R_\lambda(z,w|z',w')=\frac{(-1)^{\lambda-\bar
\lambda-\sigma_1+\bar\sigma_1+\rho_1-\bar\rho_1}
A(\lambda-\sigma_1+\rho_1)
A(\lambda-\sigma_2+\rho_2)}{[z-w]^{-\lambda+\sigma_2-\rho_1}\,[z'-w']^{-\lambda+\sigma_1-\rho_2}
[z-w']^{1+\lambda-\sigma_2+\rho_2}[z'-w]^{1+\lambda-\sigma_1+\rho_1}
}.
\end{gather*}
The straightforward  check shows that up to a prefactor this
expression coincides with the kernel for the $\CR$-operator
obtained in~\cite{SL2C}.

\looseness=1
The expression for the $\widebar{\mathcal{Q}}$ operator for $N=2$
case can be rewritten in the form
\begin{gather}
[\widebar{\mathcal{Q}}(\lambda)\Phi](z_1,\ldots,z_L)=A^L(\lambda-\sigma_1)
\prod_{k=1}^L \int d^2
\zeta_{k}[z_k-\zeta_{k+1}]^{-(1+\lambda-\sigma_{1})}
[\zeta_k-z_{k} ]^{\lambda-\sigma_2}\nonumber\\
\phantom{[\widebar{\mathcal{Q}}(\lambda)\Phi](z_1,\ldots,z_L)=}{}\times
\Phi(\zeta_1,\zeta_2,\ldots, \zeta_{L}). \label{redN=2}
\end{gather}
Again it can be checked that the expression~\eqref{redN=2}
together with equation~\eqref{QX} matches (up to a
$\lambda$-dependent prefactor) the expression for the kernel of
the Baxter $\mathcal{Q}$-operator obtained in~\cite{SL2C}.

Finally, we give one more representation for the Baxter operator
$\mathcal{Q}^{(N=2)}(\lambda)$
\begin{gather*}
\mathcal{Q}^{(N=2)}(\lambda+\sigma_2)=A^L(\lambda)\prod_{k=1}^L
\int d^2\alpha_k
[\alpha_k]^{-1-\lambda}\,[1-\alpha_k]^{\lambda-\sigma_{12}}\nonumber\\
\phantom{\mathcal{Q}^{(N=2)}(\lambda+\sigma_2)=}{}
\times\Phi((1-\alpha_1) z_L+\alpha_1 z_1, (1-\alpha_2)
z_1+\alpha_2 z_2,\ldots,
 (1-\alpha_L) z_{L-1}+\alpha_{L} z_L),
\end{gather*}
which is instructive to compare with the expression for the Baxter
$\mathcal{Q}$-operator for $su(1,1)$ spin chain~\cite{SDQ}.

\section{Summary}

In this paper we developed an approach of constructing the
solutions of the Yang--Baxter equation for the principal series
representations of $SL(N,\mathbb{C})$. We obtained the
 $\CR$-operator  as the product of  elementary operators~$\mathbb{U}_k$,
$k=1,\ldots,2N-1$. The latter, except for the
operator~$\mathbb{U}_N$, are the intertwining operators for the
principal series representations of $SL(N,\mathbb{C})$. The
opera\-tor~$\mathbb{U}_N$ is a special one. It intertwines the
tensor products of $SL(N,\mathbb{C})$ representation and a~product
of Lax operators, see equations~\eqref{SLL}, \eqref{ST}. The
operators~$\mathbb{U}_k$  satisfy the same commutation relations
as the operators of the elementary permutation, $P_k$. In other
words, they def\/ine the representation of the permutation group
$S_N$. It means that any two operators constructed from the
operator $\mathbb{U}_k$ corresponding to the same permutation are
equal. As a result a proof of the Yang--Baxter relation becomes
trivial.

We have represented the $\CR$-operator in the factorized form and
constructed the factorizing operators $\mathbb{R}^{(k)}$. Having
in mind  application of this approach to  spin chains with
generic representations of  $sl(N)$ algebra
 we f\/igured out which properties of the factorizing operators are vital to  proving of
the Yang--Baxter equation. The operators $\mathbb{R}_k$ play the
fundamental role in constructing the Baxter operators. Namely,
using them as  building blocks, one can  construct the commutative
family of the operators $\mathcal{Q}_k$ which can be identif\/ied
as the Baxter operators. We obtain the integral representation for
the latter and show that  for $N=2$ our results coincide with the
results of~\cite{SL2C}.

\subsection*{Acknowledgments}
We are grateful to M.A. Semenov-Tian-Shansky for
valuable discussions and attracting our attention to \cite{Knapp1,Knapp2}.
This work was supported by the RFFI grant 05-01-00922 and DFG
grant 436 Rus 17/9/06~(S.D.)  and by the Helmholtz Association,
contract number VH-NG-004~(A.M.)


\LastPageEnding
\end{document}